\documentclass[english]{interact}
\usepackage[T1]{fontenc}
\usepackage[latin9]{inputenc}
\usepackage{babel}
\usepackage{refstyle}
\usepackage{amsmath}
\usepackage{graphicx}
\usepackage[numbers,sort&compress]{natbib}
\usepackage[unicode=true,pdfusetitle,
 bookmarks=true,bookmarksnumbered=false,bookmarksopen=false,
 breaklinks=false,pdfborder={0 0 1},backref=false,colorlinks=false]
 {hyperref}

\makeatletter

\AtBeginDocument{\providecommand\eqref[1]{\ref{eq:#1}}}
\AtBeginDocument{\providecommand\secref[1]{\ref{sec:#1}}}
\AtBeginDocument{\providecommand\Eqref[1]{\ref{Eq:#1}}}
\AtBeginDocument{\providecommand\figref[1]{\ref{fig:#1}}}
\AtBeginDocument{\providecommand\Figref[1]{\ref{Fig:#1}}}
\RS@ifundefined{subsecref}
  {\newref{subsec}{name = \RSsectxt}}
  {}
\RS@ifundefined{thmref}
  {\def\RSthmtxt{theorem~}\newref{thm}{name = \RSthmtxt}}
  {}
\RS@ifundefined{lemref}
  {\def\RSlemtxt{lemma~}\newref{lem}{name = \RSlemtxt}}
  {}

\bibpunct[, ]{[}{]}{,}{n}{,}{,}

\makeatletter
\def\NAT@def@citea{\def\@citea{\NAT@separator}}
\makeatother

\makeatother

\begin{document}
\title{The Myth of URANS}
\author{Daniel M. Israel}
\maketitle
\begin{abstract}
Since the 1990s, RANS practitioners have observed spontaneous unsteadiness
in RANS simulations. Some have suggested deliberately using this as
a method of resolving large turbulent structures. However, to date,
no one has produced a theoretical justification for this unsteady
RANS (URANS) approach. Here, we extend the dynamical systems fixed
point analysis of \citet{speziale1989,girimaji2006b} to create a
theoretical model for URANS dynamics. The results are compared to
URANS simulations for homogeneous isotropic decaying turbulence. The
model shows that URANS can predict incorrect decay rates and that
the solution tends towards steady RANS over time. Similar analysis
for forced turbulence shows a fixed modeled energy of about 30\% of
total energy, regardless of the model parameters. The same analysis
can be used to show how hybrid type models can begin to address these
issues.
\end{abstract}

\section{\label{sec:Introduction}Introduction}

When \citet{reynolds1895} introduced the averaged equations that
bear his name, he posited an average that was local over a small area
of space or interval of time. Inspired by an analogy to kinetic theory,
he assumed that 
\begin{equation}
\overline{\bar{u}u'}=0,\label{eq:reynolds-assumption}
\end{equation}
which eliminates the cross-stress terms in the averaged equations.
For kinetic theory, this is an excellent approximation: there is an
extreme scale discrepancy between the mean-free path that characterizes
molecular motions and the scales of interest in a continuum representation
of a flow, at least for most flows of interest. For turbulence, however,
which is characterized by the broad spectrum of the energy cascade,
this assumption is wrong.

Subsequent literature has therefore dubbed any average a ``Reynolds
average'', if it has the property
\begin{equation}
\overline{f\bar{g}}=\bar{f}\bar{g},\label{eq:reynolds-average}
\end{equation}
which is a generalization of \eqref{reynolds-assumption}. Ironically,
that makes the locals spatial average Reynolds invoked not a Reynolds
average, but the equations derived in Reynolds's paper are consistent
with a Reynolds average so defined. There do exist other averages
that satisfy this property exactly, such as infinite time (often wrongly
attributed to Reynolds), homogeneous in one or more spatial directions,
or a properly constructed ensemble average.

It is evident that averages of this kind are expected to remove all
turbulent scales of motion, including the very largest eddies. The
earliest attempts to close these equations, such as \citet{taylor1915,prandtl1925},
assume that the turbulence is characterized by a single set of scales,
those corresponding to the largest, energy-containing eddies. This
class of models, which comes to be known as RANS (Reynolds-averaged
Navier-Stokes), continues to rely on the single scaling assumption.
This includes $k-\varepsilon$, $k-\omega$, and Reynolds-stress models
(RSM), all of which include one scaling equation which is used to
obtain all the remaining unclosed quantities by a dimensional scaling.
All the proposed scaling quantities (dissipation, $\varepsilon$,
inverse time scale, $\omega$, or length scale, among others) are
assumed to be algebraically related.

With the first availability of larger computers in the 1960s, starting
with the pioneering work of \citet{smagorinsky1963,lilly1967,deardorff1970},
turbulence modeling returned to the idea of a local spatial average
over regions significantly smaller than the largest turbulent eddies.
The averaging region is often identified with the computational grid
cell, although they are mathematically distinct, and this conflation
continues to cause confusion. This class of methods is called large-eddy
simulation (LES). 

The RANS literature often derives the averaged equations by invoking
a long or infinite-time average, or an ensemble average, whereas LES
papers typically formulate the filtered equations in terms of a local
spatial filter. This has lead to a widespread misunderstanding that
the choice of filter used to derive the equations is what distinguishes
RANS and LES. However, as \citet{germano1992} points out, by writing
the equations in terms of generalized moments, the moment equations
are identical for any average, provided it is linear, and commutes
with differentiation. The only difference between RANS and LES, or
between any two moment closure models, for that matter, is the value
of the unclosed subfilter terms. Since those terms are replaced with
a closure model, it should be clear that the specific average associated
with a closure is implicitly defined by the closure, rather than the
opposite. This point is made very clearly by \citet{pope2000} for
the specific case of the Smagorinsky model, and what he terms the
implied Smagorinsky filter, but the argument generalizes to any turbulence
model.

In particular, while RANS models rely on the single scaling assumption
described above, LES models invoke a cutoff scale that defines the
largest scale at which the model acts. This is typically a length
scale, and usually is related to the grid scale to assure maximum
use of the available resolution without too large a discretization
error.

It has long been known that RANS models can be used to compute time-dependent
flows in cases where the model can be considered a homogeneous spatial
average, for example, homogeneous isotropic decaying turbulence or
the temporal spreading of a mixing layer. For these examples, the
RANS equations reduce to ordinary-differential equations in the first
case and partial-differential equations in time and a single space
dimension in the second. Clearly, no turbulent structures are explicitly
resolved. Alternatively, one might consider a flow in which the time
scale of interest is very long compared to the turbulent time scale,
for example, the slow spread of a contaminant plume. In addition,
it has been common practice to solve for the time-steady RANS solution
to a problem by using a time-dependent code and iterating in time
until a steady-state is achieved.

Beginning in the 1990s, using finer grids and higher-order turbulence
models and numerical schemes, RANS practitioners began to observe
cases in which, for certain problems and models, simulations never
reached a steady state. Instead, they observed the emergence of structures
which were, at least qualitatively, similar to the large turbulent
structures observed in experiments \citep{johansson1993,durbin1995,bosch1998}.
Numerous subsequent authors have used RANS models to develop unsteady
structures deliberately, reporting mixed success \citep{travin2000,iaccarino2003,squires2005,fadai-ghotbi2008,palkin2016,pereira2019}.

The results of these and other studies sometimes look promising, but
raise numerous questions. First, and most fundamentally, it is not
clear why some models, geometries, and initial conditions lead to
unsteady solutions, while others relax to a steady solution. Second,
the reported results are mixed; some authors showing relatively good
agreement with experimental data, and other quite poor. This opens
the question whether the more successful comparisons are purely fortuitous.
At the very least, our inability to predict, a priori, when the method
will work makes it impossible to rely on for engineering prediction.
Finally, some quantities are easier to predict than others. Qualitative
agreement between visualizations of large-scale features is not surprising,
since the mechanisms that support them is often only weakly dependent
on Reynolds number. For the case of shedding behind a cylinder, for
example, \citet{pereira2019} propose that, since the shedding Strouhal
number is only weakly dependent on Reynolds number over a wide range,
it should be expected that a RANS simulation will do an adequate job
of predicting it, even if the effective viscosity is poorly predicted. 

\citet{spalart2000} and \citet{travin2004} present systematic critiques
of what they refer to as the unsteady RANS (URANS) approach. There
is some variation in terminology between various authors. Consistent
with the definitions above, we will follow the distinction made by
\citet{travin2004} that LES refers to a model which includes the
filter width as a parameter,\footnote{We can slightly generalize this to include any parameter that defines
the decomposition into resolved and modeled energy. The PANS approach\citep{girimaji2006},
for example, parameterizes the decomposition by the fraction of energy
which is modeled, rather than the cutoff length scale. } whereas RANS does not. We can also distinguish between VLES, which
is simply an LES model that resolves a smaller portion of the spectrum
than regular LES, and URANS, which is just using a traditional, unmodified
RANS model but allowing the solution to become unsteady.\footnote{This is consistent with the usage of \citet{pope2000}, but contrary
to \citet{spalart2000}, who takes URANS and VLES as synonymous.}

Although in widespread use, a physical or mathematical  justification
for URANS is missing. The underlying models typically used were all
developed for steady flows, with all scales modeled. The reason why
unsteadiness arises in certain cases, and not in others, is not understood,
other than a general idea that it is related to how unstable the flow
is, and how dissipative is the model. Physical plausibility arguments
for URANS seem to require a spectral gap between the large-structures
and the incoherent turbulent fluctuations, which is not observed in
experiment.\citep{spalart2000,travin2004} 

The inability to specify exactly what the URANS solution is supposed
to represent makes it impossible to determine if it is correct or
not. Several authors have invoked a triple-decomposition, or a phase
average, \citep{johansson1993,bosch1998,fadai-ghotbi2008,palkin2016}.
But this is an \emph{ex post facto} reinterpretation of models that
were originally derived without any accounting for large-scale resolved
structures. The models are the same whether the averages in the equations
are identified as ensemble averages or phase averages. In other words,
there is no formal mathematical justification for interpreting the
results this way. Furthermore, \citet{spalart2000} points out that
the phase-averaged structure is very different from the typical exact
structure, and the resolved field is observed to include ``incoherent''
motions. 

An alternative to URANS is to find a rational procedure for introducing
an explicit decomposition parameter into a conventional RANS model.
The flow-simulation methodology (FSM, \citep{speziale1998,israel2004,fasel2006}),
partially-averaged Navier-Stokes (PANS, \citep{girimaji2005,girimaji2006,girimaji2006b})
and partially-integrated transport model (PITM, \citep{chaouat2005,schiestel2005})
are all examples of this approach. The emergence of these methods,
which can easily be implemented in existing RANS codes, has reduced
the prevalence of URANS, but has not completely replaced it.

Without a better theoretical underpinning for the URANS approach,
it will remain in scientific limbo. A clear mathematical framework,
on the other hand, can also help us better understand URANS in comparison
with the other available modeling approaches. \citet{travin2004}
writes, ``Much work and discussion is needed to determine whether
this situation reflects a congenital flaw, or is hiding a `golden
opportunity'.'' The goal of this paper is to present a first step
in that work and discussion.

There are two primary questions to ask about URANS. First, what is
the mechanism that gives rise to unsteadiness, and when does it occur?
Second, in the absence of a parameter to set the scale for the energy
decomposition, where does that scale come from and what value does
it take? We will focus on the second question, by considering a simple
test problem, and constructing an analytic model of how the energy
decomposition evolves.

The test problem chosen is homogeneous isotropic turbulence, both
decaying and forced. This problem is not one in which URANS is typically
employed, and does not exhibit the large coherent structures seen
in such geometries as bluff body wakes. In that sense, it is a particularly
stringent test case. Nevertheless, it was selected for two reasons.
First, it is amenable to an analytic theory which can help guide our
intuitive understanding. Second, it represents the extreme in terms
of lack of scale separation; if URANS can still work in this case,
it should be able to work in cases with dynamically significant large
structures.

\section{The conventional RANS model}

We start by reviewing the conventional RANS model. In the following
discussion we restrict ourselves to the standard $k-\varepsilon$
model \citep{launder1974}, although the same analysis could easily
be extended to other RANS models. We decompose the velocity and pressure
fields into an averaged and a fluctuating part,
\begin{subequations}
\label{eq:decompose}
\begin{align}
u_{i} & =\bar{u}_{i}+u_{i}'\\
p & =\bar{p}+p'.
\end{align}
\end{subequations}
The governing equations are
\begin{subequations}
\label{eq:k-e}
\begin{gather}
\frac{\partial\bar{u}_{i}}{\partial x_{i}}=0\\
\frac{\partial\bar{u}_{i}}{\partial t}+\bar{u}_{j}\frac{\partial\bar{u}_{i}}{\partial x_{j}}=-\frac{1}{\rho}\frac{\partial\bar{p}}{\partial x_{i}}-\frac{\partial R_{ij}}{\partial x_{j}}+\nu\nabla^{2}\bar{u}_{i}\\
\frac{\partial k}{\partial t}+\bar{u}_{j}\frac{\partial k}{\partial x_{j}}=\mathcal{P}-\varepsilon+\frac{\partial}{\partial x_{j}}\left[\left(\nu+\frac{\nu_{T}}{\text{Pr}_{k}}\right)\frac{\partial k}{\partial x_{j}}\right]\\
\frac{\partial\varepsilon}{\partial t}+\bar{u}_{j}\frac{\partial\varepsilon}{\partial x_{j}}=\frac{\varepsilon}{k}\left(C_{\varepsilon1}\mathcal{P}-C_{\varepsilon2}\varepsilon\right)+\frac{\partial}{\partial x_{j}}\left[\left(\nu+\frac{\nu_{T}}{\text{Pr}_{\varepsilon}}\right)\frac{\partial\varepsilon}{\partial x_{j}}\right].\label{eq:k-e-dissipation}
\end{gather}
\end{subequations}
 The Reynolds stress is defined as
\begin{equation}
R_{ij}=\overline{u_{i}'u_{j}'},
\end{equation}
and modeled with a linear eddy-viscosity
\begin{equation}
R_{ij}=-2\nu_{T}\bar{S}_{ij}+\frac{2}{3}k\delta_{ij},
\end{equation}
with
\begin{equation}
\bar{S}_{ij}=\frac{1}{2}\left(\frac{\partial\bar{u}_{i}}{\partial x_{j}}+\frac{\partial\bar{u}_{j}}{\partial x_{i}}\right).
\end{equation}
The eddy-viscosity is
\begin{equation}
\nu_{T}=C_{\mu}\frac{k^{2}}{\varepsilon},
\end{equation}
and the production of turbulent kinetic energy is
\begin{equation}
\mathcal{P}=-\frac{\partial\bar{u}_{i}}{\partial x_{j}}R_{ij}=2\nu_{T}\bar{S}_{ij}\bar{S}_{ij}.
\end{equation}

The over-bar represents an average, which is often taken to be a homogenous
spatial average, for problems with one or more homogeneous spatial
dimensions, or a time average, for steady problems. For more general
problems, it is common to invoke an ensemble average. These averages
are often equivalent, assuming ergodicity holds. The turbulent kinetic
energy, 
\begin{equation}
k=\frac{1}{2}\overline{u_{i}'u_{i}'},\label{eq:tke}
\end{equation}
and the dissipation rate, 
\begin{equation}
\varepsilon=\nu\overline{\frac{\partial u_{i}'}{\partial x_{j}}\frac{\partial u_{i}'}{\partial x_{j}}},\label{eq:dissipation}
\end{equation}
are then the energy and dissipation in the scales of motion too small
to be resolved in the $\bar{u}_{i}$ velocity field.

In fact, the exact averaged equations, which these equations are a
model for, are identical, regardless of which average is invoked.
This implies that the average that is obtained from an actual simulation
is a consequence of the specific model closure used, and is independent
of the average specified by the author, if any. In the case of the
$k-\varepsilon$ model, the average is implied by the modeling assumption
that the eddy-viscosity can be scaled with a velocity characteristic
of the entire turbulent motion, $k^{1/2}$, and a length-scale
\[
L\sim\frac{k^{3/2}}{\varepsilon},
\]
as well as the rescalings that go into modeling the dissipation rate.
Consistent with our definitions in \secref{Introduction}, the RANS
has no explicit parameter defining a partition of resolved and modeled
scales. An alternative closure approach would be to explicitly introduce
such a parameter to obtain an LES model. For example, if we impose
a length scale for computing the eddy-viscosity, $L\sim C_{S}\Delta x$,
then the equations are immediately closed without the need for a dissipation
rate \eqref{k-e-dissipation}, and we recover the one-equation LES
model of \citet{schumann1975,yoshizawa1982}.

Particular care must be given to initial and boundary conditions.
\Eqref[s]{k-e} allow for injecting both a modeled and a resolved
component of turbulent fluctuations through the initial and boundary
conditions. There is nothing in the equations, however, which ensures
that this injection is consistent with the energy decomposition implied
by the model. In fact, it is fair to ask whether the possibility of
forcing the model in an inconsistent manner constitutes a defect in
the model, or just a caveat on how we treat initial and boundary conditions.

Applying \eqref[s]{k-e} to homogeneous isotropic decaying turbulence,
using the conventional RANS approach, we assume that the average removes
all the turbulent scales. Since the flow is statistically homogeneous,
all the average quantities are uniform in space, and the equations
reduce to ordinary differential equations. To indicate this, we will
replace the over-bar with an angle bracket. This could represent a
spatial average over all of space, or an ergodically equivalent, suitably
constructed, ensemble average. The average velocities are identically
zero, $\left\langle u_{i}\right\rangle =0$, and the $k-\varepsilon$
equations reduce to a set of ordinary differential equations,
\begin{subequations}
\label{eq:k-e-hit}
\begin{align}
\frac{d\left\langle k\right\rangle }{dt} & =-\left\langle \varepsilon\right\rangle \label{eq:hit-tke}\\
\frac{d\left\langle \varepsilon\right\rangle }{dt} & =-C_{\varepsilon2}\frac{\left\langle \varepsilon\right\rangle ^{2}}{\left\langle k\right\rangle }.\label{eq:hit-dissipation}
\end{align}
\end{subequations}
The solution to this problem is well known to be
\begin{equation}
\left\langle k\right\rangle =k_{0}\left(\frac{t}{t_{0}}\right)^{-n},
\end{equation}
where the decay rate is 
\begin{equation}
n=\frac{1}{C_{\varepsilon2}-1}.
\end{equation}
For this simple RANS model, the decay rate is a constant. Data from
experiments and simulations suggest that this may not be quite right,
but for a particular experiment, over a wide range of Reynolds number,
it is a reasonable approximation \citep{mohamed1990,lavoie2007}.

\section{Unsteady RANS}

We can also try to simulate this problem using a URANS. This means
we use the same $k-\varepsilon$ RANS equations (\ref{eq:k-e}), but
now we include all the unsteady and locally varying terms. For this
problem, unsteadiness will not arise spontaneously, so we need to
introduce it through the choice of initial conditions. We wish do
to so in a way that is consistent with the idea that the initial conditions
represent some suitably averaged real turbulent state. In practice,
we can achieve this by computing a fully resolved turbulent field
using direct-numerical simulation of the unfiltered Navier-Stokes
equations. The simulation is allowed to evolve until the flow reaches
the asymptotic state of homogeneous decay. We then choose a low-pass
filter with a specified length-scale. The velocities can be filtered
directly, and the turbulence quantities can be obtained using equations
(\ref{eq:tke}) and (\ref{eq:dissipation}).

The governing equations are now
\begin{subequations}
\label{eq:urans}
\begin{gather}
\frac{\partial\bar{u}_{i}}{\partial x_{i}}=0\\
\frac{\partial\bar{u}_{i}}{\partial t}+\bar{u}_{j}\frac{\partial\bar{u}_{i}}{\partial x_{j}}=-\frac{1}{\rho}\frac{\partial\bar{p}}{\partial x_{i}}+\frac{\partial}{\partial x_{j}}\left[\left(\nu+\nu_{T}^{<}\right)\left(\frac{\partial\bar{u}_{i}}{\partial x_{j}}+\frac{\partial\bar{u}_{j}}{\partial x_{i}}\right)\right]\label{eq:urans-momentum}\\
\frac{\partial k_{<}}{\partial t}+\bar{u}_{j}\frac{\partial k_{<}}{\partial x_{j}}=\mathcal{P}_{<}-\varepsilon_{<}+\frac{\partial}{\partial x_{j}}\left[\left(\nu+\frac{\nu_{T}^{<}}{\text{Pr}_{k}^{<}}\right)\frac{\partial k_{<}}{\partial x_{j}}\right]\label{eq:urans-k-sfs}\\
\frac{\partial\varepsilon_{<}}{\partial t}+\bar{u}_{j}\frac{\partial\varepsilon_{<}}{\partial x_{j}}=\frac{\varepsilon_{<}}{k_{<}}\left(C_{\varepsilon1}^{<}\mathcal{P}_{<}-C_{\varepsilon2}^{<}\varepsilon_{<}\right)+\frac{\partial}{\partial x_{j}}\left[\left(\nu+\frac{\nu_{T}^{<}}{\text{Pr}_{\varepsilon}^{<}}\right)\frac{\partial\varepsilon_{<}}{\partial x_{j}}\right],\label{eq:urans-eps-sfs}
\end{gather}
with
\[
\mathcal{P}_{<}=2\nu_{T}^{<}\bar{S}_{ij}\bar{S}_{ij},
\]
and
\[
\nu_{T}^{<}=C_{\mu}\frac{k_{<}^{2}}{\varepsilon_{<}}.
\]
\end{subequations}
Note, these equations are formally identical to the RANS equations
(\ref{eq:k-e}), however, we have introduced super- or subscript less-than
decorations to various quantities to emphasize that they now represent
subfilter quantities. Both the velocities and the turbulence quantities
now vary in both time and space.

The bar now represents a local \emph{filter}, which does not remove
all turbulent scales, and the term \emph{average}, denoted by angle
brackets will be reserved for the homogeneous average over the entire
box. It is important to note that, as mentioned above, the filter
corresponding to a specific model is implicitly defined by the model,
and is therefore not known in a closed form that can be used to compute
the initial conditions. We will assume that the effect of using slightly
different low-pass filters leads to, at most, minimal, short-lived
transients as the flow adjusts.

Unlike LES models, which have an explicit cut-off length scale specified,
when using a conventional RANS model to try to compute unsteadiness,
the model does not include a parameter that controls which scales,
or how much of the energy, is to be captured by the model. This makes
it difficult to say what constitutes the correct answer. Instead we
will focus on two questions. How does the partition of energy between
resolved and model scales evolve? And, how well does the URANS predict
the overall decay rate? The first question addresses how useful the
model is: given that we do not control the energy partition, does
it behave in a manner that produces an informative result? The second
question is more about correctness. There may not be a clear right
answer for each component, resolved and modeled, of the energy, but
the total energy should give something like a power-law decay, as
observed in experiments, with at most a slowly varying decay rate.
In particular, if the energy partition changes, it should do so in
a way that does not degrade the solution for the total energy.

In order to explore these questions, it would be nice if we could
reduce equations (\ref{eq:urans}) to a set of ordinary differential
equations. We could then employ the dynamical systems and fixed point
analysis first introduced by \citet{speziale1989} in the steady RANS
context, and applied to URANS by \citet{girimaji2006b}. In their
approach, they omitted the unsteady transport term as not contributing
to the global energy balance. We take a slightly different approach
and formally average the unsteady equations. To do this, we first
introduce the decomposition
\[
k=\left\langle k_{>}\right\rangle +\left\langle k_{<}\right\rangle ,
\]
where
\begin{subequations}
\label{eq:k-decomposition}
\begin{gather}
k_{<}=\frac{1}{2}\left(\overline{u_{i}u_{i}}-\bar{u}_{i}\bar{u}_{i}\right)\\
k_{>}=\frac{1}{2}\left(\bar{u}_{i}\bar{u}_{i}-\left\langle \bar{u}_{i}\right\rangle \left\langle \bar{u}_{i}\right\rangle \right),
\end{gather}
\end{subequations}
are the energy of the subfilter and resolved scales, respectively.
This decomposition requires the assumption that $\left\langle \bar{f}\right\rangle \approx\left\langle f\right\rangle $.
This can be demonstrated explicitly for certain filters and averages.
A similar decomposition exists for the dissipation.

Taking the average of \eqref{urans-k-sfs} and using homogeneity,
the equation for the subfilter energy is exactly
\begin{equation}
\frac{\partial\left\langle k_{<}\right\rangle }{\partial t}=\left\langle \mathcal{P}_{<}\right\rangle -\left\langle \varepsilon_{<}\right\rangle .\label{eq:sfs-energy}
\end{equation}
The equation for the average resolved energy can be derived by multiplying
\eqref{urans-momentum} by $\bar{u}_{i}$, and averaging,
\begin{equation}
\frac{\partial\left\langle k_{>}\right\rangle }{\partial t}=-\left\langle \mathcal{P}_{<}\right\rangle -\left\langle \varepsilon_{>}\right\rangle .\label{eq:resolved-energy}
\end{equation}
The subfilter dissipation \eqref{urans-eps-sfs}, when averaged, yields
\begin{equation}
\frac{\partial\left\langle \varepsilon_{<}\right\rangle }{\partial t}=C_{\varepsilon1}^{<}\left\langle \frac{\varepsilon_{<}}{k_{<}}\mathcal{P}_{<}\right\rangle -C_{\varepsilon2}^{<}\left\langle \frac{\varepsilon_{<}^{2}}{k_{<}}\right\rangle .\label{eq:sfs-dissipation}
\end{equation}
This is an exact equation for how the subfilter dissipation rate will
evolve under the action of the RANS model given by \eqref{k-e}. In
order to solve this equation, an additional level of closure modeling
is required, since the two averaged terms on the right-hand side are
not closed. That is, this equations includes a closure model relative
to the bar filter, but still requires closing relative to the bracket
filter. The simplest model for \eqref{sfs-dissipation} is
\begin{gather}
\left\langle \frac{\varepsilon_{<}^{2}}{k_{<}}\right\rangle \approx\frac{\left\langle \varepsilon_{<}\right\rangle ^{2}}{\left\langle k_{<}\right\rangle }\\
\left\langle \frac{\varepsilon_{<}}{k_{<}}\mathcal{P}_{<}\right\rangle \approx\frac{\left\langle \varepsilon_{<}\right\rangle }{\left\langle k_{<}\right\rangle }\left\langle \mathcal{P}_{<}\right\rangle .
\end{gather}
This is correct to second order in the fluctuations \citep{haering2022}.
For the remaining unclosed quantities, we use
\begin{gather}
\left\langle \mathcal{P}_{<}\right\rangle \approx\left\langle \nu_{T}\right\rangle \left\langle \bar{\omega}^{2}\right\rangle \label{eq:production-model}\\
\left\langle \varepsilon_{>}\right\rangle =\nu\left\langle \bar{\omega}^{2}\right\rangle ,\label{eq:dissipation-model}
\end{gather}
where we also decompose the average eddy-viscosity as 
\begin{equation}
\left\langle \nu_{T}\right\rangle =C_{\mu}\frac{\left\langle k_{<}\right\rangle ^{2}}{\left\langle \varepsilon_{<}\right\rangle }.
\end{equation}
Note that \eqref{dissipation-model} is exact, whereas \eqref{production-model}
neglects the spatial variation of the eddy-viscosity. 

We can now close the three \eqref[s]{sfs-energy,resolved-energy,sfs-dissipation}
by adding an evolution equation for filtered enstrophy magnitude,
$\left\langle \bar{\omega}^{2}\right\rangle $. To obtain this we
start by noting that the dissipation \eqref{hit-dissipation} is essentially
an enstrophy equation,
\begin{equation}
\frac{d\left\langle \omega^{2}\right\rangle }{dt}=-C_{\varepsilon2}\frac{\nu\left\langle \omega^{2}\right\rangle ^{2}}{\left\langle k\right\rangle }.
\end{equation}
If we assume that the dynamics of a URANS simulation can be modeled
as a Navier-Stokes simulation with a reduced viscosity of $\nu_{T}$,
rather than $\nu$, then a plausible model for the enstrophy of the
resolved scales is
\begin{equation}
\frac{\partial\left\langle \bar{\omega}^{2}\right\rangle }{\partial t}=-C_{\varepsilon2}^{>}\frac{\left\langle \nu_{T}\right\rangle \left\langle \bar{\omega}^{2}\right\rangle ^{2}}{\left\langle k_{>}\right\rangle }.
\end{equation}
(A similar argument can be found in \citep{pope2000} for deriving
an estimate of the filtered spectrum produced by an LES model.)

The coefficient $C_{\varepsilon2}^{>}$ plays the same role for the
enstrophy dissipation of the filtered scales as the $C_{\varepsilon2}^{<}$
plays for the subfilter scales in the original RANS model, and is
the only parameter of the reduced order model. The parameters $C_{\varepsilon1}^{<}$
and $C_{\varepsilon2}^{<}$ are not tunable parameters of the reduced
order model, they must be set to whatever the values are used in the
URANS simulations we are comparing to. For the results presented here,
the value used is the standard $C_{\varepsilon2}$ parameter setting,
$C_{\varepsilon2}^{>}=1.92$.

The complete set of modeled equations is now
\begin{subequations}
\label{eq:ode-model}
\begin{gather}
\frac{\partial\left\langle k_{>}\right\rangle }{\partial t}=-\left\langle \mathcal{P}_{<}\right\rangle -\left\langle \varepsilon_{>}\right\rangle \label{eq:ode-resolved-tke}\\
\frac{\partial\left\langle \bar{\omega}^{2}\right\rangle }{\partial t}=-C_{\varepsilon2}^{>}\frac{\left\langle \nu_{T}\right\rangle \left\langle \bar{\omega}^{2}\right\rangle ^{2}}{\left\langle k_{>}\right\rangle }\label{eq:ode-enstrophy}\\
\frac{\partial\left\langle k_{<}\right\rangle }{\partial t}=\left\langle \mathcal{P}_{<}\right\rangle -\left\langle \varepsilon_{<}\right\rangle \label{eq:ode-sfs-tke}\\
\frac{\partial\left\langle \varepsilon_{<}\right\rangle }{\partial t}=\frac{\left\langle \varepsilon_{<}\right\rangle }{\left\langle k_{<}\right\rangle }\left(C_{\varepsilon1}^{<}\left\langle \mathcal{P}_{<}\right\rangle -C_{\varepsilon2}^{<}\left\langle \varepsilon_{<}\right\rangle \right).\label{eq:ode-sfs-dissipation}
\end{gather}
\end{subequations}
The first thing to note about these equations is that if we sum \eqref{ode-resolved-tke,ode-sfs-tke},
we recover the exact total energy \eqref{hit-tke}. (This is, in fact,
a property of the exact unclosed equations, so will hold for any closure
model.) It is evident that the subfilter production term, $\left\langle \mathcal{P}_{<}\right\rangle $
represents the transfer of energy from the resolved to the subfilter
scales. As such, it is the primary term which sets the decomposition
scale. This point is widely misunderstood, and bears emphasis. \emph{The
decomposition into resolved and subfilter components is not unique,
and the partition of energy is set implicitly by the model through
the action of $\left\langle \mathcal{P}_{<}\right\rangle $.}

Summing \eqref{ode-sfs-dissipation} and viscosity times (\ref{eq:ode-enstrophy})
does not recover the total dissipation \eqref{hit-dissipation}, however.
Instead we obtain
\begin{equation}
\frac{\partial\varepsilon}{\partial t}=\left(C_{\varepsilon1}^{<}\frac{\left\langle \varepsilon_{<}\right\rangle }{\left\langle k_{<}\right\rangle }-C_{\varepsilon2}^{>}\frac{\left\langle \varepsilon_{>}\right\rangle }{\left\langle k_{>}\right\rangle }\right)\left\langle \mathcal{P}_{<}\right\rangle -C_{\varepsilon2}^{<}\frac{\left\langle \varepsilon_{<}\right\rangle ^{2}}{\left\langle k_{<}\right\rangle }
\end{equation}
If we assume that the corresponding terms play similar roles in the
dissipation equations as they do in the kinetic energy equations,
then the first two term represent the dissipation of enstrophy out
of the large scales, and into the small scales, and should therefore
cancel. The last term should represent the dissipation of small scale
dissipation, and should equal the right-hand side of \eqref{hit-dissipation}.
That this is not the case allows our model for the URANS to explore
deviations from the RANS power law decay. This is in distinction from
the analysis of \citet{girimaji2006b}, which assumes the total dissipation
still obeys \eqref{hit-dissipation}.

Rather than analyzing the full set of equations (\ref{eq:ode-model}),
we can construct a reduced order model in terms of the primary quantities
of interest. The first of these is the energy partition, defined as
\begin{subequations}
\begin{equation}
f=\frac{\left\langle k_{<}\right\rangle }{k}.
\end{equation}
The equation for this quantity can be written in terms of two other
quantities, the subfilter production to dissipation ratio, and the
subfilter Reynolds number,
\begin{gather}
g=\frac{\left\langle \mathcal{P}_{<}\right\rangle }{\left\langle \varepsilon_{<}\right\rangle }\\
R=\frac{\left\langle \nu_{T}\right\rangle }{\nu},
\end{gather}
\end{subequations}
to form a closed set,
\begin{subequations}
\label{eq:rom}
\begin{gather}
\frac{df}{dt^{*}}=\left(R^{-1}fg+f+g-1\right)f\\
\frac{dg}{dt^{*}}=2\left(C_{\varepsilon2}^{<}-1\right)g-\left(C_{\varepsilon2}^{>}\frac{f}{1-f}+2\left(C_{\varepsilon1}^{<}-1\right)\right)g^{2}\label{eq:g}\\
\frac{dR}{dt^{*}}=\left(\left(2-C_{\varepsilon1}^{<}\right)g+C_{\varepsilon2}^{<}-2\right)R.
\end{gather}
\end{subequations}
To fully non-dimensionalize the system, we have introduced the non-dimensional
time
\begin{equation}
\frac{dt}{dt^{*}}=\frac{\left\langle k_{<}\right\rangle }{\left\langle \varepsilon_{<}\right\rangle }.
\end{equation}

This model has a singularity at when the subfilter Reynolds number
goes to zero, but for even moderate Reynolds number, the singular
term is negligible, and the $f$ and $g$ equations evolve independently
of $R$. We can better understand the behavior of $R$ by noting that
\[
R=C_{\mu}\frac{g+R}{R}f^{2}\text{Re}_{t},
\]
where the turbulent Reynolds number is defined as,
\begin{equation}
\mathrm{Re}_{t}=\frac{k^{2}}{\nu\varepsilon}.
\end{equation}
Therefore the singular term behaves as
\[
R^{-1}fg=C_{\mu}^{-1}\frac{R}{g+R}\frac{g}{f}\text{Re}_{t}^{-1}.
\]
This term will blow up in two situations. The first is when $\text{Re}_{t}\rightarrow0$,
that is, when the turbulence is almost completely dissipated. We can
neglect this case as uninteresting to our current investigation. The
second is when $f\rightarrow0$, which is the DNS limit, when the
flow is almost completely resolved. For this latter situation, for
this term to be $\mathcal{O}\left(1\right)$, we must have $f\sim\text{Re}_{t}^{-1}$.
This case can also be ignored, since if $f$ was that small, we would
just run DNS instead of URANS. Consequently, the reduced-order model
can be well approximated as
\begin{subequations}
\label{eq:rom-hi-R}
\begin{gather}
\frac{df}{dt^{*}}=\left(f+g-1\right)f\\
\frac{dg}{dt^{*}}=2\left(C_{\varepsilon2}^{<}-1\right)g-\left(C_{\varepsilon2}^{>}\frac{f}{1-f}+2\left(C_{\varepsilon1}^{<}-1\right)\right)g^{2}.
\end{gather}
\end{subequations}
The same equations can be obtained directly by neglecting the viscous
dissipation in the resolved scales, $\left\langle \varepsilon_{>}\right\rangle $,
in \eqref{ode-resolved-tke}.

\begin{figure}
\begin{centering}
\includegraphics[width=0.8\textwidth]{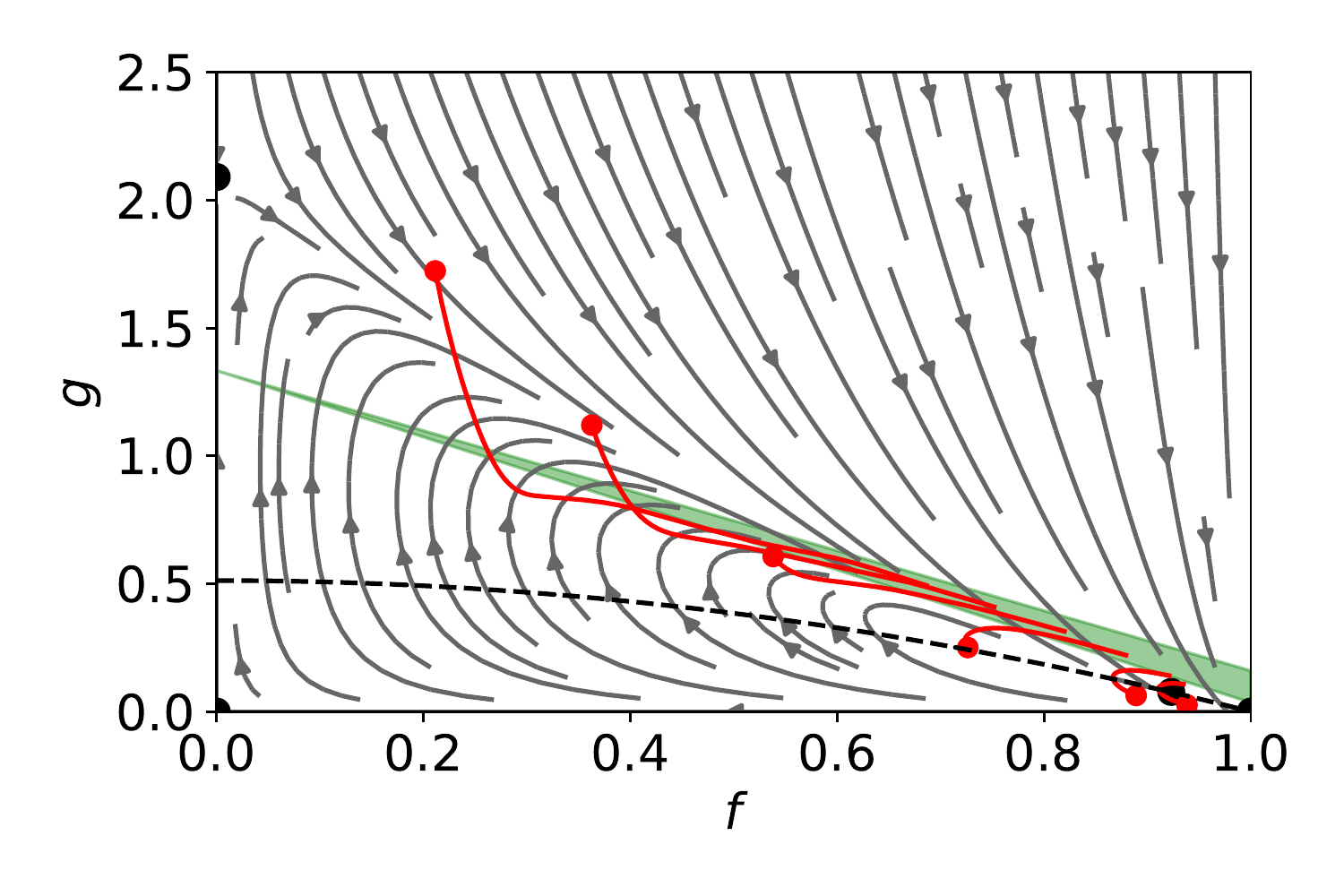}
\par\end{centering}
\caption{\label{fig:rom}Phase space evolution of \eqref{rom-hi-R}. The grey
trajectories are the reduced order model, with the fixed points in
black. Red lines are trajectories from simulation data (with points
indicating the beginning of each trajectory). The green shaded region
is where $n=1.15-1.45$. The dashed curve is \eqref{ic-phase-space}.}

\end{figure}

This model has four fixed points,
\begin{subequations}
\label{eq:fixed-points}
\begin{align}
f & =0 & g & =0\\
f & =0 & g & =\frac{C_{\varepsilon2}^{<}-1}{C_{\varepsilon1}^{<}-1}\\
f & =1 & g & =0\\
f & =\frac{2C_{\varepsilon1}^{<}-2C_{\varepsilon2}^{<}}{2C_{\varepsilon1}^{<}-C_{\varepsilon2}^{>}-2} & g & =\frac{2C_{\varepsilon2}^{<}-C_{\varepsilon2}^{>}-2}{2C_{\varepsilon1}^{<}-C_{\varepsilon2}^{>}-2}.\label{eq:attracting-fp}
\end{align}
\end{subequations}
The last one is an attractor that governs the behavior of the system.
As can be seen in \figref{rom}, all trajectories tend towards this
fixed point, which is very near the RANS limit.

For comparison, \figref{rom} also shows several real trajectories
from numerical experiments, in red. To obtain these, a fully resolved
DNS simulation for forced homogeneous turbulence was run until it
reached equilibrium. The velocity fields obtained were filtered (as
described above) to obtain initial conditions for the URANS. Each
trajectory was created using a different filter width to generate
the initial condition. The trajectory was then computed using a URANS.
Details of the simulations can be found in \citep{towery2022}.

What the figure clearly shows is that, regardless of the initial condition,
the URANS evolves towards something very like a steady RANS, with
the resolved fluctuations decaying away. The model suggests that the
URANS will not go all the way to the RANS limit of $f=1$, however
the data is not sufficient to determine if this is the case. What
is clear, is that the user has no control over the range of scales
which are physically resolved. However, it should be noted that the
rate at which the URANS relaxes towards RANS is extremely slow in
turbulence time scales, which may be one reason why some URANS practitioners
are left with the impression that the initial choice of energy partition
persists.

Having considered the evolution of the energy partition, we now examine
the decay rate of the total turbulent kinetic energy. For a flow in
which the decay rate may be changing, there are several ways to measure
this \citep{perot2006}. We choose
\begin{equation}
n=\left(\frac{d}{dt}\left[\frac{k}{\varepsilon}\right]\right)^{-1},
\end{equation}
which has the advantage that it can be computed from \eqref[s]{ode-model}
in terms of our reduced order model,
\begin{equation}
n=\frac{f}{C_{\varepsilon2}^{<}-C_{\varepsilon1}^{<}g-f}.\label{eq:decay-rate}
\end{equation}
On our trajectory plot, curves of constant $n$ are straight lines
through the point $f=0$, $g=C_{\varepsilon2}^{<}/C_{\varepsilon1}^{<}$.
The green shaded region in \figref{rom} is where the decay matches
the observed range of experimental data, $n=1.15-1.45$.

It is clear that the shaded region, where the decay rate matches experiment,
occupies only a small fraction of phase space. Whether this is a problem
for the URANS approach in practice depends largely on how much time
the URANS spends outside this small region. The plotted trajectories
suggest that the URANS does evolve to this region, given sufficient
time. Another question might be, where do typical initial conditions
fall relative to this region. For example, given a local spatial filter,
characterized by a cutoff length-scale, we could filter DNS data at
various cutoffs, to obtain an initial condition for each one. Each
initial condition would define one point in phase space, and the points
for all possible filter widths would define a curve. We can approximate
that curve using a stick-spectrum analysis, to obtain the dashed curve,
\eqref{ic-phase-space}, shown in \figref{rom}. (The details of the
derivation are given in \secref{Stick-spectrum-analysis}). From this
analysis, it appears that the initial conditions thus produced fall
well below the shaded region, meaning their will be a significant
adjustment before the correct energy decay rate is observed. The red
trajectories, on the other hand, start well above (at least for small
$f$), the difference probably being a result of the specific way
they are generated, starting with a forced turbulence and only later
being allowed to decay, as described in \citep{towery2022}.

An interesting feature of the reduced-order model is that if we use
\eqref{attracting-fp} and \eqref{decay-rate} to compute the decay
rate at the fixed point, we get
\begin{equation}
n=\frac{2}{C_{\varepsilon2}^{>}}\label{eq:decay-fp}
\end{equation}
In other words, the decay rate is independent of the underlying RANS
model, and is a function of the dynamics of the resolved scales alone.
The model coefficient we chose predicts a decay rate of $n\approx1.02$,
noticeably lower than might be expected from experiments. However,
from \eqref{decay-rate} we see that changing $C_{\varepsilon2}^{>}$
does not actually change the decay rate at any point in phase space,
rather it moves the fixed point itself. In fact, lowering $C_{\varepsilon2}^{>}$
much beyond the value adopted here moves the fixed point outside of
the allowable range $0\leq f\leq1$ , so that the $f=1$ fixed point
becomes the attractor.

A similar factor explains why we cannot adjust the model to lower
the fixed point on the upper-left, to try to improve the fit. The
location of this point is controlled by the URANS model coefficients
$C_{\varepsilon1}^{<},C_{\varepsilon2}^{<}$, which are not adjustable
parameters of the reduced-order model, and must be set to match the
coefficients settings used in the URANS we are comparing to.

\begin{figure}
\begin{centering}
\includegraphics[width=0.8\textwidth]{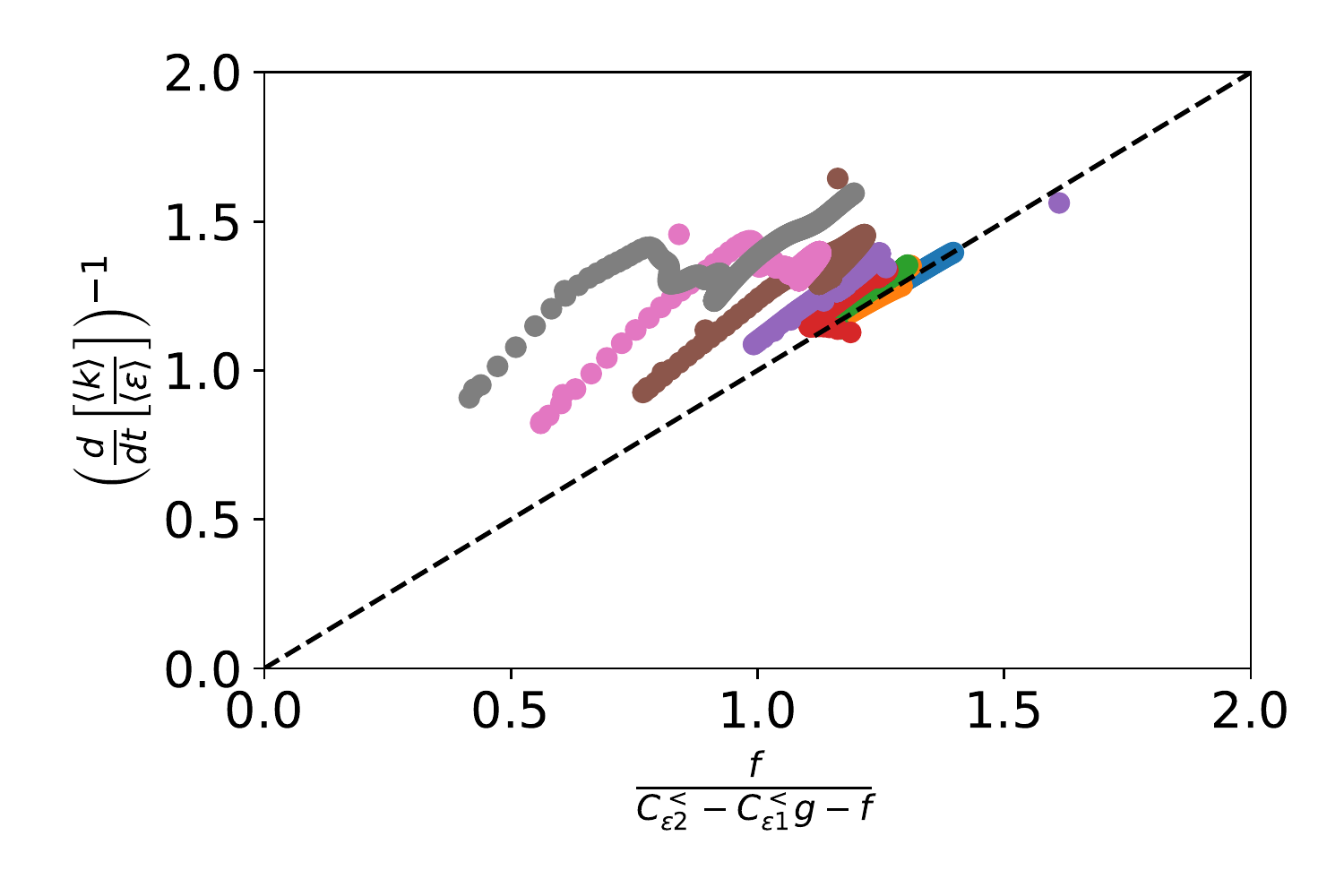}
\par\end{centering}
\caption{\label{fig:decay-rate}The measured decay rate compared versus the
decay rate predicted by the model (equation \ref{eq:decay-rate})
for the URANS simulations \citep{towery2022}.}

\end{figure}

Another way we can assess the accuracy of the model is to compare
the measured growth rate in the simulations to the model prediction
using \eqref{decay-rate} with $f$ and $g$ computed from the simulation
data. This is shown in \figref{decay-rate}. If the model was perfect,
all the data would fall on the dashed line. As is clear from the plot,
for some of the trajectories (specifically, the ones with higher $f$),
the agreement is excellent. For the trajectories closer to DNS, the
agreement is not as good; the relationship is still linear, but the
measured decay rate is about 30\% high. Also, for those trajectories
for which we observe a kink in the track in \figref{rom}, the scaling
relationship in \figref{decay-rate} seems to be different before
and after the kink.

Overall, the reduced order model \eqref{rom-hi-R} gives us a good
feel for the dynamics of the URANS. We can see that the solution can
deviate substantially from the correct decay rate as it evolves. However,
there appears to be a preferred trajectory, a roughly straight path
between two of the fixed points (on the upper-left and lower-right
in \figref{rom}), along the lower portion of which the decay rate
is close to the experimental values. This may give a false sense that
the URANS approach is correct. At least from the reduced-order model,
it appears that the preferred trajectory is not aligned with a contour
of constant decay rate, although the full simulation trajectories
are less clear.

\section{Forced Turbulence}

Decaying turbulence is arguably not analogous to problems, such as
bluff body wakes, in which unsteadiness arises spontaneously in URANS.
It is more similar to problems where the initial condition includes
a resolved perturbation or geometric feature that creates an initial
large-scale structure which decays over long times, such as\citep{morgan2016,haines2013}.
For the wake, where the flow is globally unstable, the instability
acts as a kind of forcing term which sustains the unsteadiness in
the URANS solution. The equivalent mechanism for homogeneous turbulence
would be forced turbulence, which we consider in this section.

Consider a DNS of homogeneous isotropic turbulence generated by a
large scale forcing. The flow evolves under a statistically stationary
stochastic forcing until statistical equilibrium is reached. At this
point, as for the decaying problem, the flow is decomposed into resolved
and subfilter quantities using a specified cutoff length-scale. The
problem is then restarted, using the URANS equations, initialized
with the resolved and modeled fields, and the evolution is continued
with the same forcing. This is again repeated with a series of different
cutoffs.

\begin{figure}
\begin{centering}
\includegraphics[width=0.8\textwidth]{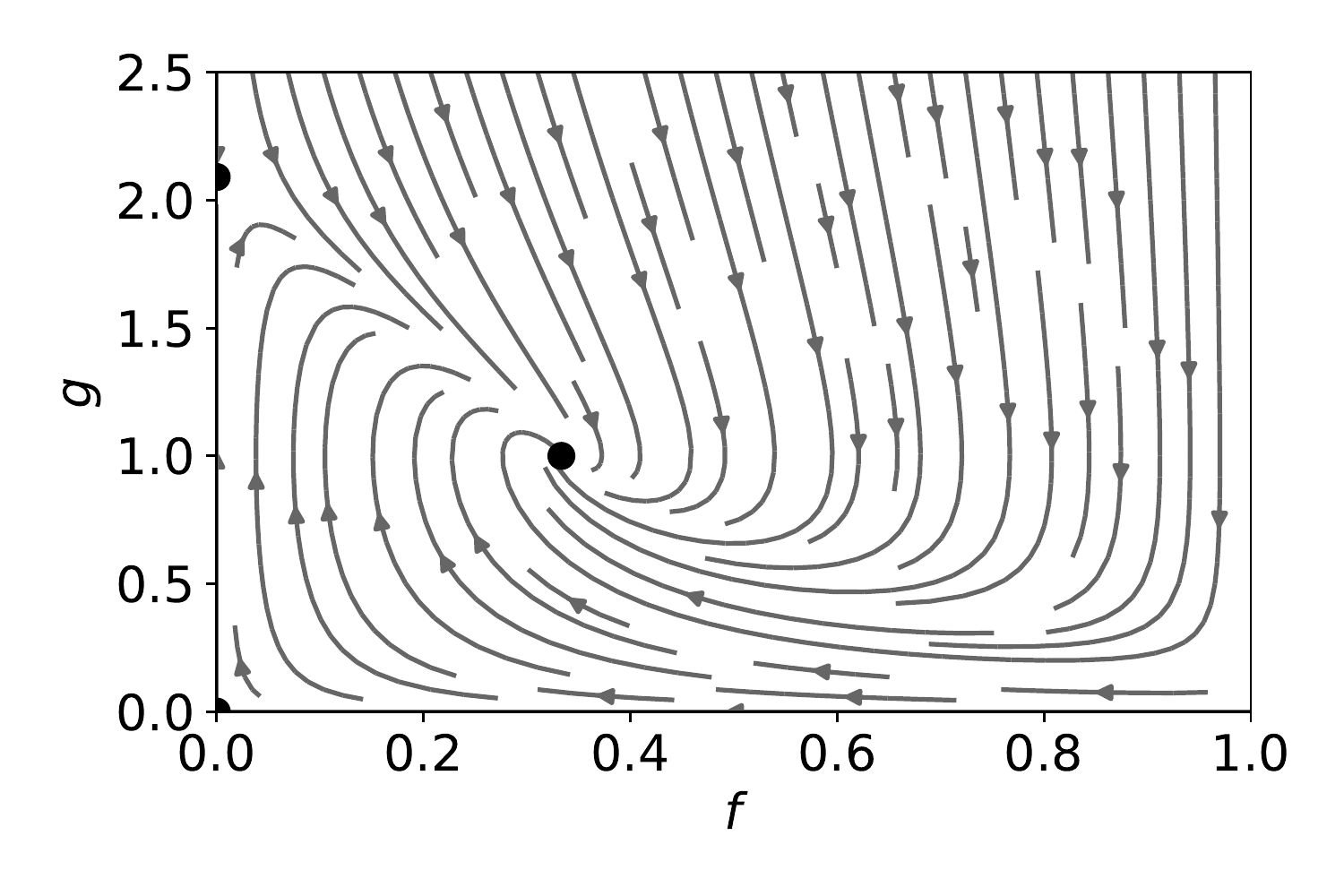}
\par\end{centering}
\caption{\label{fig:rom-forced}Phase space evolution for homogeneous forced
turbulence, \eqref{rom-forced,g}. The grey trajectories are the reduced
order model, with the fixed points in black.}
\end{figure}

To model this problem, we will make the following assumptions. First,
that the forcing is entirely in the resolved component, regardless
of the cutoff length-scale. Second, that the flow remains in equilibrium,
so that we can assume that the forcing is always balanced by the dissipation,
$\mathcal{F}=\varepsilon$. Finally, that the forcing does not directly
effect the dissipation or enstrophy. In that case, \eqref{ode-enstrophy,ode-sfs-tke,ode-sfs-dissipation}
remain unchanged, however \eqref{ode-resolved-tke} becomes
\[
\frac{\partial\left\langle k_{>}\right\rangle }{\partial t}=-\left\langle \mathcal{P}_{<}\right\rangle -\left\langle \varepsilon_{>}\right\rangle +\mathcal{F},
\]
and
\[
\frac{dk}{dt}=0.
\]
The reduced-order model for the forced case is, therefore
\begin{gather}
\frac{df}{dt^{*}}=f\left(g-1\right),\label{eq:rom-forced}
\end{gather}
and the $g$ equation is still \eqref{g}. The fixed points are
\begin{subequations}
\label{eq:fp-forced}
\begin{align}
f & =0 & g & =0\\
f & =0 & g & =\frac{C_{\varepsilon2}^{<}-1}{C_{\varepsilon1}^{<}-1}\\
f & =\frac{2\left(C_{\varepsilon1}^{<}-C_{\varepsilon2}^{<}\right)}{2C_{\varepsilon1}^{<}-2C_{\varepsilon2}^{<}-C_{\varepsilon2}^{>}} & g & =1
\end{align}
\end{subequations}
With the standard $k-\varepsilon$ coefficients, this results in an
attracting fixed point at $g=1$ and $f\approx0.3$, as can be seen
in \figref{rom-forced}.

Unlike the unforced case, the energy partition does not trend towards
a steady RANS solution in the forced case. Instead, the forcing acts
as a continuous source of resolved scale energy. The fixed point represents
the balance between the dissipation of resolved scale motion due to
the model and the production of resolved scale motion due to the forcing.
A value of $g=1$ at the fixed point means that the energy input due
to forcing is the same as the transfer (production) between resolved
and subfilter scales, which is the same as the subfilter dissipation,
that is, a classic cascade process. The fixed points we obtain are
different from those in \citep{girimaji2006b} because we do not assume
that the total dissipation rate follows the steady RANS equation,
as we describe above.

\section{Implications for RANS based hybrid models}

The fundamental problem with the URANS approach is that, regardless
of how the simulation is initialized, conventional RANS model equations
are designed to account for all the turbulent scales of motion. When
used to simulate only a range of scales, the dynamics of the solution
fundamentally disagrees with the modeling assumptions. In order to
fix this problem, there exists a class of hybrid turbulence models
which, through various mathematical procedures, attempt to rescale
the subfilter quantities in the RANS equations to properly account
for the partition between resolved and modeled scales. These include
the flow-simulation methodology (FSM) \citep{speziale1998,fasel2006},
partially averaged Navier-Stokes (PANS) \citep{girimaji2006,girimaji2006b},
and the partially integrated transport model (PITM) \citep{schiestel2005}.
The reduced-order model approach from the previous section is easily
extended to these hybrid models. In fact, for models where the only
change is a modification to the model coefficients, the reduced-order
model (\ref{eq:rom}) still applies, it only needs to be replotted
with the appropriate coefficient values.

To demonstrate this, we consider the PANS model. In the PANS approach,
the $C_{\varepsilon2}^{<}$ coefficient is replaced with a rescaled
version,
\begin{equation}
C_{\varepsilon2}^{*}=f_{k}\left(C_{\varepsilon2}^{<}-C_{\varepsilon1}^{<}\right)+C_{\varepsilon1}^{<},\label{eq:pans}
\end{equation}
$f_{k}$ is a model parameter that is supposed to set the ratio of
the modeled to total energy. This is distinguished from $f$, which
is the actual ratio achieved by the simulation. In other words, $f_{k}$
is an input to the model, and $f$ is a diagnostic; if the PANS model
worked exactly as intended, $f_{k}$ and $f$ would be equal. 

\begin{figure}
\begin{centering}
\includegraphics[width=0.8\textwidth]{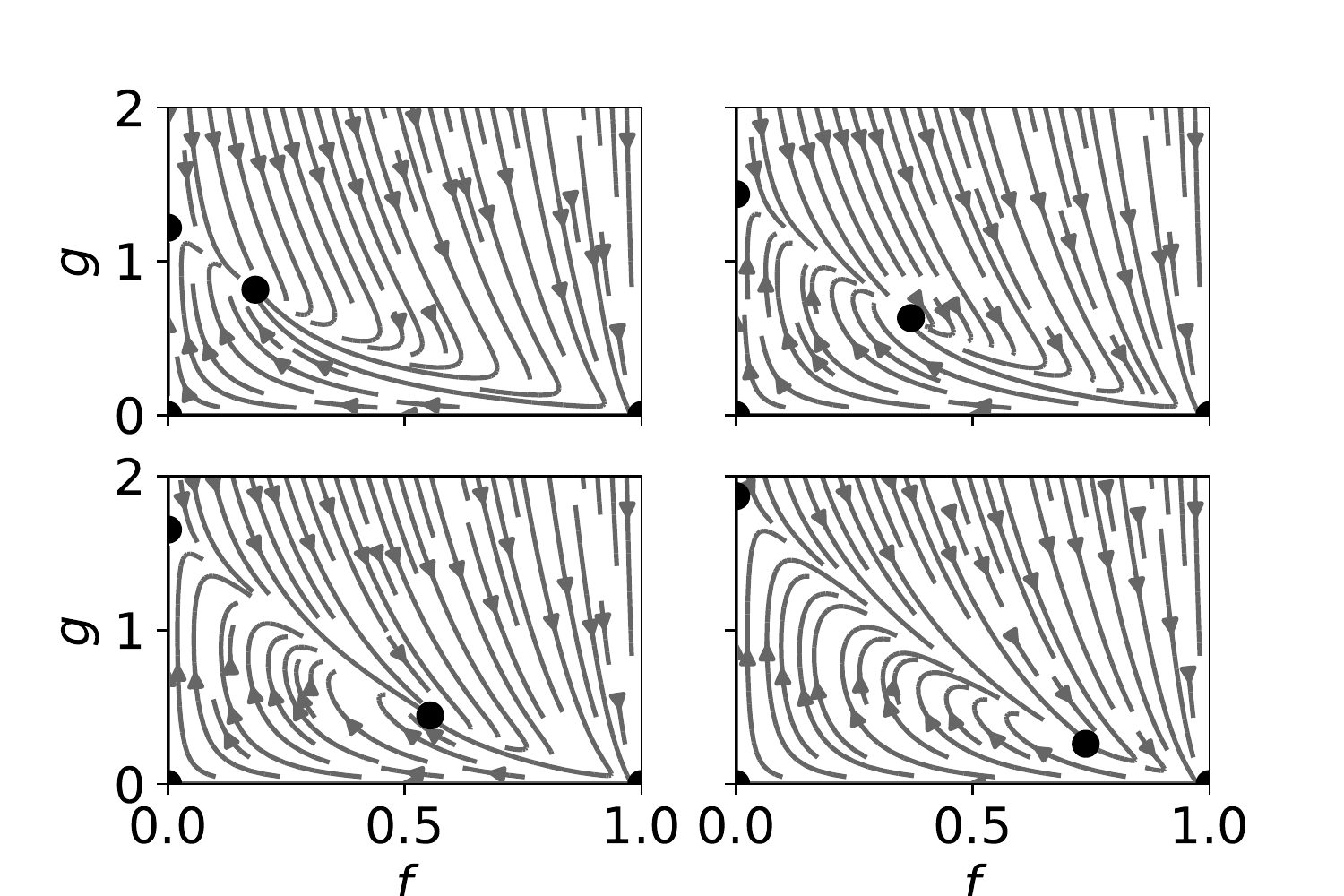}
\par\end{centering}
\caption{\label{fig:pans}Phase space evolution of unforced  PANS with different
values of $f_{k}$. From top left, $f_{k}=0.2,0.4,0.6,0.8$.}
\end{figure}
\Figref{pans} shows the trajectory maps for the PANS models with
four different values of $f_{k}$. The fixed points move depending
on $f_{k}$, and the behavior is as desired: as $f_{k}$ decreases,
the value of $f$ at the attracting fixed point also decreases. We
can substitute $C_{\varepsilon2}^{*}$ from (\ref{eq:pans}) for $C_{\varepsilon2}^{<}$
in the expression for the reduced-order model fixed point value of
$f$, (\ref{eq:attracting-fp}), to obtain an analytic relationship
between $f_{k}$ and $f$ at the fixed point,
\begin{equation}
f=\frac{2\left(C_{\varepsilon1}^{<}-C_{\varepsilon2}^{<}\right)}{2C_{\varepsilon1}^{<}-C_{\varepsilon2}^{>}-2}f_{k}.\label{eq:pans-f}
\end{equation}
For the standard values of the model coefficients, this is approximately
$f\approx0.96f_{k}$, which is very close to the desired equality.
Note also, that from \eqref{decay-fp}, the growth rate at the fixed
point is independent of $f_{k}$.

\begin{figure}
\begin{centering}
\includegraphics[width=0.8\textwidth]{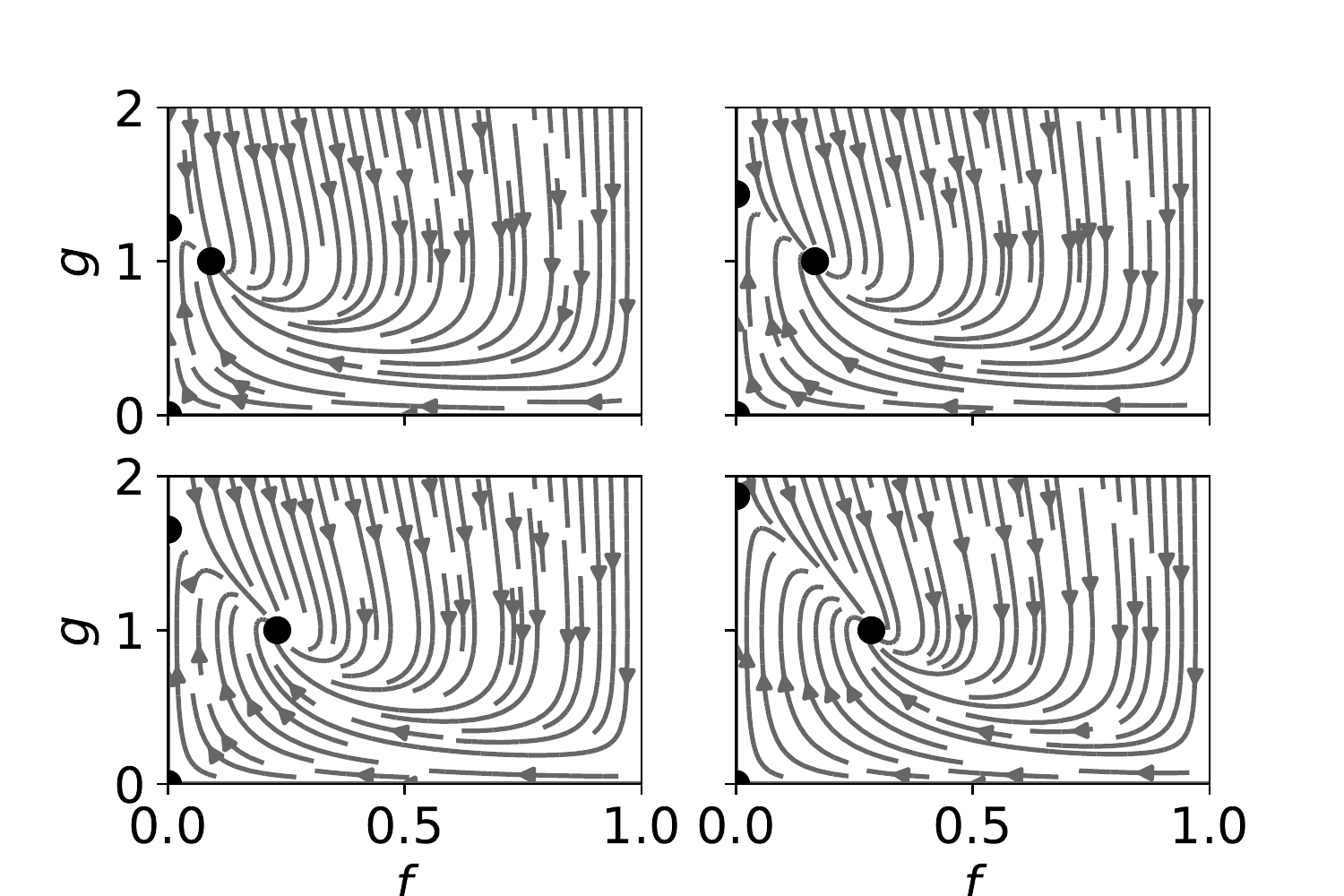}
\par\end{centering}
\caption{\label{fig:pans-forced}Phase space evolution for forced PANS with
different values of $f_{k}$. From top left, $f_{k}=0.2,0.4,0.6,0.8$.}
\end{figure}

We can conduct the same experiment for the forced case. In this case,
we do not expect the result to relax to a steady solution in the limit
of $f_{k}\rightarrow1$, since the underlying RANS model does not
do so. Instead, the level of unsteadiness is restricted to range of
$f\approx0-0.3$, as can be seen in \figref{pans-forced}.

In other words, for the unforced case, the PANS model does exactly
what it is designed to do. Unlike URANS, where the limiting behavior
relaxes to a conventional RANS, the PANS model tends towards a value
of $f$ very close to the value of the input parameter $f_{k}$. Nevertheless,
it is important to be aware that the PANS model also can have significant
deviations from the fixed point, particularly if the initial condition
is not chosen so that both $f$ and $g$ are at the fixed point value
for the specified $f_{k}$. For the forced case, the relationship
between the input $f_{k}$ and the observed $f$ is more complex.
Obviously, the PANS model cannot enforce a steady result at the $f_{k}=1$
limit if the underlying RANS model cannot.

\section{Conclusions}

Conventional RANS models were developed using a single-scale assumption.
Such closures are only justified where any resolved unsteadiness exhibits
a ``spectral gap,'' or, equivalently, the large scales, such as
the total turbulent kinetic energy and the integral length scale,
completely characterize the turbulence. Even for problems with large
unsteady turbulent structures, these models were designed to produce
the correct solution for the time mean, when solved for the steady-state
solution (i.e., when the time-derivative terms are set to zero). Proponents
of the URANS approach have not produced a rigorous justification as
to why these models sometimes give rise to unsteady solutions if allowed
to evolve in time, nor what precisely those unsteady solutions represent.
Without clearly identifying what the models should do, it is hard
to say definitely whether the results they do produce are correct
or not, or even what would be the correct way to employ them.

One key limitation in our understanding is that existing research
into URANS consists of case studies. Even very careful analysis can
lead to uncertain, and even contradictory, conclusions, given the
enormous parameter space of all potential applications of URANS. What
has been missing is a theoretical framework for understanding what
URANS actually does. This paper presents a first step in that direction.

The proposed model can be viewed as an extension of earlier fixed
point analyses. The power of the approach lies in viewing the evolution
of the URANS simulation as a dynamical system, with the phase space
dynamics giving insight into the global behavior of the model. The
purpose of the model is not to be an accurate predictive tool, but
rather to reproduce the qualitative behaviors we expect to see in
URANS simulations. Comparisons to actual simulation data indicates
that the model does just that.

Without a clear metric for correctness, it is impossible to say definitively
that URANS is wrong, but these results show that the idea that a steady
RANS model can simply by applied to unsteadiness and a correct answer
obtained is a myth. Defenders of the URANS approach might try to argue
that the results in \figref{rom}, showing that the URANS has a preference
for trajectories that fall near the region of correct decay rate,
vindicates the approach. However, there are three serious difficulties
with this view. First, the fact that a URANS solution will decay in
time until almost all resolved scales are dissipated and a steady
RANS solution is achieved, is neither a useful model behavior, nor
consistent with most users expectations for an unsteady model. Second,
both the stick spectrum analysis and the simulation data initialized
with forced turbulence show that initializing the URANS so that it
starts in a state where the decay rate is reasonable is not a trivial
undertaking. Finally, while the model may, in certain cases, predict
the decay rate properly, in others it can undergo extreme transients
into highly non-physical behaviors.

Extending the analysis to the hybrid PANS model, we see that for decaying
turbulence PANS solves the first of these issues, placing the fixed
point at exactly the resolution the user requested. However, the path
to the fixed point, and, in particular, inconsistent initial conditions,
still are issues that require further study. Here too, the forced
case show more complicated dynamics. This is consistent with the observation
that the hybrid model cannot be calibrated to enforce an energy partition
more coarse than the underlying RANS model upon which it is built.

The approach does need to be extended to more common applications
of URANS. Work is underway for an extension to free shear and mixing
layers. In the meantime, the case of forced turbulence provides a
possible analog for the mechanisms at work in URANS of globally unstable
flows, with the forcing acting as a proxy for the global instability.
These results show that in such a case the forcing can keep the unsteadiness
alive. The energy partition is controlled by a balance between the
model damping and the forcing input.

In either case, further analysis would be served by the proponents
of the URANS approach clearly defining metrics for what they believe
the correct behavior of the model should be. In the meantime, users
of models would be advised to have a very clear understanding of what
can and cannot be expected of RANS models before applying them to
unsteady problems.

\section*{Acknowledgements}

The author would like to express his gratitude to Dr. Colin Towery
for providing simulation data for comparison, as well as very helpful
discussions. Thanks also to Dr. Filipe Pereira for his insightful
comments on the manuscript. This research was funded by the LANL Mix
and Burn project under the DOE ASC, Physics and Engineering Models
program.

\appendix

\section{\label{sec:Stick-spectrum-analysis}Stick spectrum analysis}

The initial conditions for a URANS type simulation (or, equivalently,
for a PANS/hybrid/SRS simulation) cannot fall in any arbitrary point
in $f,g,R$ space. The production-to-dissipation ratio, $g$, will
depend on the energy partition, $f$. Another way to look at it is,
given a particular filter that is parameterized by a length scale,
$\ell$, for each $\ell$ we will have specific values of $f$, $g$,
and $R$. That is, $f=f\left(\ell\right)$, $g=g\left(\ell\right)$,
and $R=R\left(\ell\right)$ are all functions of $\ell$. Choosing
a different filter shape will result in a slightly different curve,
but it is likely that this will be a weak effect.

We can estimate this dependence by using an assumed spectrum analysis
\citep[and references therein]{ristorcelli2006}. If we take $L$
as the characteristic length scale of the large scales, and the Kolmogorov
scale, $\eta$, as characterizing the smallest scales, we can use
an assumed spectrum of
\begin{equation}
E\left(\kappa\right)=\begin{cases}
C\left\langle \varepsilon\right\rangle ^{2/3}\kappa^{-5/3} & \kappa\in\left[2\pi/L,2\pi/\eta\right]\\
0, & \text{elsewhere}
\end{cases}.\label{eq:assumed-spectrum}
\end{equation}
The subfilter turbulent kinetic energy is
\begin{equation}
\left\langle k_{<}\right\rangle =\int_{\frac{2\pi}{\ell}}^{\infty}E\left(\kappa\right)d\kappa.
\end{equation}
Integrating the assumed spectra \eqref{assumed-spectrum} gives an
estimate of
\begin{equation}
\left\langle k_{<}\right\rangle =\frac{3}{2}C\left\langle \varepsilon\right\rangle ^{2/3}\left(\frac{\ell}{2\pi}\right)^{2/3}\left[1-\left(\frac{\eta}{\ell}\right)^{2/3}\right].\label{eq:tke-sgs-estimate}
\end{equation}
The total turbulent kinetic energy is just the RANS limit of \eqref{tke-sgs-estimate}:
when $\ell$ goes to infinity (actually $\ell=L$) all scales are
captured, so
\begin{equation}
\left\langle k\right\rangle =\frac{3}{2}C\left\langle \varepsilon\right\rangle ^{2/3}\left(\frac{L}{2\pi}\right)^{2/3}\left[1-\left(\frac{\eta}{L}\right)^{2/3}\right].\label{eq:tke-estimate}
\end{equation}
Using this we can obtain an estimate for $f$,
\begin{equation}
f=\left(\frac{\ell}{L}\right)^{2/3}\frac{\left[1-\left(\frac{\eta}{\ell}\right)^{2/3}\right]}{\left[1-\left(\frac{\eta}{L}\right)^{2/3}\right]}.
\end{equation}

To compute $g$ and $R$, we need the subfilter enstrophy, 
\begin{equation}
\left\langle \omega^{\prime2}\right\rangle =\int_{\frac{2\pi}{\ell}}^{\infty}\kappa^{2}E\left(\kappa\right)d\kappa,
\end{equation}
we can use our assumed spectrum to obtain
\begin{equation}
\left\langle \omega^{\prime2}\right\rangle =\frac{3}{4}C\left\langle \varepsilon\right\rangle ^{2/3}\left(\frac{2\pi}{\eta}\right)^{4/3}\left[1-\left(\frac{\eta}{\ell}\right)^{4/3}\right],\label{eq:enstrophy-estimate}
\end{equation}
and, therefore,
\begin{equation}
R=3C_{\mu}C\frac{\left\langle \varepsilon\right\rangle ^{2/3}}{\nu^{2}}\left(\frac{\ell}{2\pi}\right)^{4/3}\left(\frac{\eta}{2\pi}\right)^{4/3}\frac{\left[1-\left(\frac{\eta}{\ell}\right)^{2/3}\right]^{2}}{\left[1-\left(\frac{\eta}{\ell}\right)^{4/3}\right]}.
\end{equation}
To find the leading order scaling of $R$, we can eliminate $\varepsilon$
and $\nu$ in terms of $L$ and $\eta$. To do this, we cannot use
the usual dimensional estimates, which are not correct for our stick
spectrum. Instead, we must use \eqref{tke-estimate} and
\begin{equation}
\left\langle \varepsilon\right\rangle =\nu\left\langle \omega^{2}\right\rangle =\nu\lim_{\ell\rightarrow\infty}\left\langle \omega_{<}^{2}\right\rangle =\frac{3}{4}C\nu\left\langle \varepsilon\right\rangle ^{2/3}\left(\frac{2\pi}{\eta}\right)^{4/3}\left[1-\left(\frac{\eta}{L}\right)^{4/3}\right].
\end{equation}
With these we find that $R$ scales with $\text{Re}_{t}$,
\begin{equation}
R=C_{\mu}\text{Re}_{t}\left(\frac{\ell}{L}\right)^{4/3}\frac{\left[1-\left(\frac{\eta}{L}\right)^{4/3}\right]}{\left[1-\left(\frac{\eta}{\ell}\right)^{4/3}\right]}\frac{\left[1-\left(\frac{\eta}{\ell}\right)^{2/3}\right]^{2}}{\left[1-\left(\frac{\eta}{L}\right)^{2/3}\right]^{2}}.
\end{equation}

Finally, 
\begin{align}
g & =\frac{\left\langle \mathcal{P}_{<}\right\rangle }{\left\langle \varepsilon_{<}\right\rangle }\approx\frac{\left\langle \nu_{T}^{<}\right\rangle \left\langle \bar{\omega}^{2}\right\rangle }{\nu\left\langle \omega_{<}^{2}\right\rangle }=R\frac{\left\langle \bar{\omega}^{2}\right\rangle }{\left\langle \omega_{<}^{2}\right\rangle },
\end{align}
or
\[
g=\frac{27}{16}C_{\mu}C^{3}\left(1-\left(\frac{\ell}{L}\right)^{4/3}\right)\left[1-\left(\frac{\eta}{L}\right)^{4/3}\right]^{2}\frac{\left[1-\left(\frac{\eta}{\ell}\right)^{2/3}\right]^{2}}{\left[1-\left(\frac{\eta}{\ell}\right)^{4/3}\right]^{2}}.
\]

If we further consider the case when $\eta\rightarrow0$, then
\begin{gather*}
f=\left(\frac{\ell}{L}\right)^{2/3}\\
g=\frac{27}{16}C_{\mu}C^{3}\left(1-\left(\frac{\ell}{L}\right)^{4/3}\right),
\end{gather*}
or
\begin{equation}
g=\frac{27}{16}C_{\mu}C^{3}\left(1-f^{2}\right).\label{eq:ic-phase-space}
\end{equation}

\bibliographystyle{unsrtnat}
\bibliography{abbrev,library,refs,books}

\end{document}